\documentclass[aps,prc,twocolumn,showpacs,amsmath,amssymb,nofootinbib]{revtex4-2}
\usepackage{graphicx}  
\usepackage{color}
\usepackage{times}
\usepackage{inputenc}
\usepackage{float}
\usepackage{bm}
\usepackage{ulem}
\usepackage{multirow}
\usepackage{float}
\usepackage{url}
\usepackage{natbib}
\usepackage{mathrsfs}
\usepackage{physics}
\usepackage{comment}
\usepackage[table,xcdraw]{xcolor}
\usepackage{booktabs,makecell}
\usepackage{comment}
\usepackage{amsmath}
\usepackage{pifont}
\usepackage[modulo]{lineno}
\setpagewiselinenumbers
\modulolinenumbers[1]

\usepackage[colorlinks=true,citecolor=blue,urlcolor=blue,linkcolor=blue]{hyperref}
\begin{document}
\title{{\it f}-mode oscillations of dark matter admixed quarkyonic neutron star}
\author{D. Dey$^{1,2}$}
\author{Jeet Amrit Pattnaik$^3$}
\email{jeetamritboudh@gmail.com }
\author{R. N. Panda$^{3}$}
\author{M. Bhuyan$^1$}
\author{S. K. Patra$^{3}$}
\email{debabrat.d@iopb.res.in }

\affiliation{\it $^{1}$Institute of Physics, Sachivalaya Marg, Bhubaneswar-751005, India}
\affiliation{\it $^{2}$Homi Bhabha National Institute, Training School Complex, 
Anushakti Nagar, Mumbai 400094, India}
\affiliation{\it $^{3}$Department of Physics, Siksha $'O'$ Anusandhan, Deemed to be University, Bhubaneswar -751030, India}
\date{\today}
\begin{abstract}
We systematically investigate $f-$mode oscillations ($\ell$ = 2) in quarkyonic neutron stars with dark matter, employing the Cowling approximation within the framework of linearized general relativity. The relativistic mean-field approach is used to compute various macroscopic properties of neutron stars. The analysis focuses on three key free parameters in the model: transition density, QCD confinement scale, and dark matter (DM) Fermi momentum, all of which significantly affect the properties of $f-$mode oscillations. The inclusion of dark matter in quarkyonic equations of state leads to notable variations in $f-$mode frequencies. Despite these changes, several universal relations among the oscillation properties are found to hold, demonstrating their robustness in the presence of dark matter.
\end{abstract}
\maketitle
\section{Introduction}
\label{intro}
A neutron star is born as a remnant of the core-collapse supernova explosion. The core of these objects contains some of the most compact forms of matter known to exist in the observable universe. The central density can exceed 5 to 10 times the nuclear saturation density, which is $\sim$0.16$\rm fm^{-3}$. The mass of such a star is  $\sim$ 2 $M_\odot$ and the radius is typically $\sim$ 12 km. Unlike normal stars, the general theory of relativity plays a significant role in the formation of these relativistic compact objects. Although the microscopic constituents of NS are still an open question, mostly it comprises neutrons, a few fractions of protons as well as leptons. Due to its extremely high-density environment, there may exist many exotic particles such as free quarks, condensed pions, and kaons, which play a pivotal role in shaping the macroscopic behaviour of NS \cite{doi:10.1126/science.1090720, Burrows2000}. On the observational front, a plethora of new measurements of masses and radii are presented to the scientific community, which combined with theoretical study can put forward order meaningful constraints and explore the nature of the equation of state along with micromacro aspects of NS.  Recently, gravitational wave measurement GW170817 of the binary NS merger \cite{PhysRevLett.119.161101, PhysRevLett.121.091102, PhysRevLett.121.161101, Capano2020} along with NICER X-ray observations of PSR J0030+0451 \cite{Riley_2019, Miller_2019}  has provided the constraint that $1.4 \ M_\odot$ stars have radii $R_{1.4}\leq 13.5$ km. The nature of dense is again challenged by the measurement of NS masses greater than or equal to $2 \ M_\odot$ \cite{doi:10.1146/annurev-nucl-102711-095018,2016ARA&A..54..401O}.
Mclerran and Reddy, to explain these findings, proposed a model dubbed quarkyonic matter. This model assumes that both nucleons and quarks can be effectively treated as quasi-particles which is manifested due to the cross-over transition among nucleons and quarks at the neutron star core owing to its high density. The model also impacts some of the early models of quark hadron phase transition such as the MIT bag model \cite{BURGIO200219} and Nambu-Jona-Lasinio model \cite{PhysRevC.60.025801}. The quarkyonic matter model provides a first-order phase transition where quarks drip out of nucleons occupying the lower Fermi momentum states, while nucleons themselves occupy higher Fermi momenta states. This results in a swift increase in pressure near transition density that can be seen in the non-monotonic behaviour in the speed of sound which respects the large density asymptotic behaviour. The components of this model consist of up (u) and down (d) quarks and neutrons which are later on extended by Jhao and Lattimer \cite{PhysRevD.102.023021} to meet specific conditions of steller matter that includes beta equilibrium between leptons and hadrons, charge neutrality condition and chemical equilibrium among quarks and nucleons. They have included two nucleon species as well as other types of leptons in their approach.
In addition, we include the possibility of dark matter (DM) inside the NS, where the effective field theory-motivated relativistic mean field (E-RMF) approach is considered \cite{dey2024}. Neutron star environments are ideal for capturing and containing the DM inside them due to their immense gravitational potential. The nature of DM is still shrouded in mystery and many theoretical models exist to date to describe the various observational effects of dark matter. Several candidates exist such as sterile neutrinos, bosonic DM,  weakly interacting massive particles (WIMPs), neutrino, axions, feebly interacting massive particles, etc. \cite{PhysRevD.83.083512, DM2, 2017IJMPA..3230023B, Hall2010, PhysRevD.69.035001,2014JHEP...08..093H, PhysRevD.99.043016, Duffy_2009}.  DM inside can significantly alter the macroscopic properties of NS and several theoretical studies were conducted \cite{DM5, DM6, DM7, DM8, DM9, DM10, DM11, DM12, DM13, routray}. Many studies suggest that DM nucleon scattering leads to the accumulation of DM inside NS. By transferring the kinetic energy of DM the scattering process can warm the NS to near-infrared regime \cite{dark_matter_3, dark_matter_4}. There exist several direct and indirect detection experimental approaches to find pieces of evidence for the existence of DM.  \cite{Bernabei2008, Bernabei2010, PhysRevLett.101.091301, doi:10.1126/science.1186112, conrad2014indirect}. Our DM model in this work is Neutralino which is a fermion that interacts with nucleonic matter via the Higgs portal mechanism  \cite{DM1, DM10}.
The quantity known as the equation of state (EOS) is of fundamental importance when it comes to studying the densest states of matter found inside the interior of NS. Several non-relativistic and relativistic theoretical frameworks have been developed in order to determine the EOS. \cite{doi:10.1080/14786435608238186, SKYRME1958615, PhysRevC.5.626, CHABANAT1998231, PhysRevC.58.220, STONE2007587, PhysRevC.85.035201, PhysRevC.21.1568}. Among these Relativistic Mean Field Theory (RMF) stands out to be a prominent approach as the theory quite successfully describes various properties of finite nuclei over the nuclear landscape. It is also versatile since its region of applicability ranges from finite nuclei to superheavy and exotic nuclei to NS matter \cite{PhysRevC.63.044303, PhysRevC.97.045806, patt1, patt2, patt3, patt4,patt5}. By formulation, It is covariant in its structure based on relativistic quantum field theory which was developed to describe nuclear many-body problem. It also adheres to the method of self-consistency to find the relevant equation of motion. Here, we employ the RMF model with an appropriate parameter set to investigate the impact of DM on quarkyonic stars.
Traditionally Neutron star observation spans a significant area of the electromagnetic spectrum ranging from radio to X-rays to gamma rays with the space-based and ground-based telescopes. But in addition to the electromagnetic regime, with the success of the detection of gravitation waves, tighter constraints of observable has been achieved. NS can emit a large amount of gravitational wave if it oscillates or a merger happens between NS- NS or NS-BH. Among many detections, Recent observations of GW170817 (NS-NS collision) as well as  GW200105  snd  GW200115 (NS-BH mergers) by LIGO-VIRGO-KAGRA observatories have ushered a new age of multi-messenger astronomy \cite{GW170817, GW200105/115}. The oscillation of Neutron stars is quite an interesting problem by itself since these carry the nature of internal constituents and viscous forces that damp these modes\cite{Kokkotas}. Various oscillation modes can exist inside an NS such as fundamental $f$-mode, $p$-mode, $g$-mode, $r$-mode, $w$-modes etc. These odes are categorized according to the restoring force that brings the perturbed star to the equilibrium state. For example, $f$-, $p$-, $g$- modes are driven primarily by pressure and buoyancy forces respectively, r modes are triggered by Coriolis force while the g modes are activated during merger or inspiral event of binary NS. \cite{gmode1, gmode2, gmode3} . Since g modes can probe deep inside the NS it is very sensitive to the internal composition of NS. However, The signature of gravitational waves is too weak to be observed by current-generation observatories.

Our focus in this work is on the $f$-mode oscillations of neutron stars, which are studied extensively by Andersson and Kokkotas \cite{Anderson}, establishing a relationship between various macroscopic properties of neutron stars and the $f$-mode frequency as well as its damping times. Since these oscillation equations are complicated in nature, often background metric perturbations are neglected. This kind of approximation known as the Cowling approximation provides good and reasonable estimates of the frequency although a fully general relativistic approach is required more a better computation  \cite{Cowling1, cowling01, cowling02}. We compute  $f$-modes of DM admixed quarkyonic stars using Cowling approximation. This paper is organized in the following way: In Sec. \ref{formalism},
we discuss the theoretical formalism of the nuclear model, quarkyonic model, dark matter model, equilibrium state, and f-mode oscillation equations of NS. In Sec. \ref{results}, we discuss the results of our work followed by a conclusion in Sec. \ref{Conclusions}.

\section{Theoretical framework}
\label{formalism}
\subsection{Nuclear Model}
\label{a}
The success of RMF formalism is evident from its capacity to describe infinite nuclear matter, finite nuclei as well as  NS matter. Its range of applicability is quite large i.e., sub-saturation nuclear matter to supra-saturation matter properties. This model can also be applied from the stability line to the limit of exotic nuclei. These features of RMF make it a suitable and robust framework for the superdense state of matter inside NS. The Lagrangian density is constructed to represent various interactions among nucleons via mesons including its self and cross-couplings. Around 200 parameter sets are developed within the framework of RMF by fitting with different experimental and empirical data\cite{PhysRevC.55.540,PhysRevC.63.044303,PhysRevC.74.045806,PhysRevC.70.058801,LALAZISSIS200936,PhysRevC.82.055803,PhysRevC.82.025203,PhysRevC.84.054309,PhysRevC.85.024302,PhysRevC.97.045806,PhysRevC.102.065805}. For the present study we select the effective RMF (E-RMF) model \cite{E-RMF1, E-RMF2, E-RMF3, E-RMF4, E-RMF5}. Here the Lagrangian density has  $4^{\rm th}$ order cross and self-coupling. While the kinetic term of leptons is also there. Therefore, the energy density ($\mathcal{E}_{\rm NML}$) and pressure ($P_{\rm NML}$)  can be calculated from the stress-energy tensor for the system of nuclear matter with leptons are given as \cite{E-RMF5}.

\begin{eqnarray}
\label{eq:eden}
{\cal E}_{\rm NML} & = & \sum_{i=p,n} \frac{g_s}{(2\pi)^{3}}\int_{0}^{k_{f_{i}}} d^{3}k\, \sqrt{k^{2} + M_{\rm nucl.}^{*2}}\nonumber\\
&&
+n_{b} g_\omega\,\omega+m_{\sigma}^2{\sigma}^2\Bigg(\frac{1}{2}+\frac{\kappa_{3}}{3!}\frac{g_\sigma\sigma}{M_{\rm nucl.}}+\frac{\kappa_4}{4!}\frac{g_\sigma^2\sigma^2}{M_{\rm nucl.}^2}\Bigg)
\nonumber\\
&&
 -\frac{1}{4!}\zeta_{0}\,{g_{\omega}^2}\,\omega^4
 -\frac{1}{2}m_{\omega}^2\,\omega^2\Bigg(1+\eta_{1}\frac{g_\sigma\sigma}{M_{\rm nucl.}}+\frac{\eta_{2}}{2}\frac{g_\sigma^2\sigma^2}{M_{\rm nucl.}^2}\Bigg)
 \nonumber\\
&&
 + \frac{1}{2} (n_{n} - n_{p}) \,g_\rho\,\rho
 -\frac{1}{2}\Bigg(1+\frac{\eta_{\rho}g_\sigma\sigma}{M_{\rm nucl.}}\Bigg)m_{\rho}^2
 \nonumber\\
 && 
-\Lambda_{\omega}\, g_\rho^2\, g_\omega^2\, \rho^2\, \omega^2
+\frac{1}{2}m_{\delta}^2\, \delta^{2}
\nonumber\\
 && 
+\sum_{j=e,\mu}  \frac{g_s}{(2\pi)^{3}}\int_{0}^{k_{f_{j}}} \sqrt{k^2 + m^2_{j}} \, d^{3}k,
\end{eqnarray}
and
\begin{eqnarray}
\label{eq:press}
P_{\rm NML} & = & \sum_{i=p,n} \frac{g_s}{3 (2\pi)^{3}}\int_{0}^{k_{f_{i}}} d^{3}k\, \frac{k^2}{\sqrt{k^{2} + M_{\rm nucl.}^{*2}}} \nonumber\\
&& - m_{\sigma}^2{\sigma}^2\Bigg(\frac{1}{2} + \frac{\kappa_{3}}{3!}\frac{g_\sigma\sigma}{M_{\rm nucl.}} + \frac{\kappa_4}{4!}\frac{g_\sigma^2\sigma^2}{M_{\rm nucl.}^2}\Bigg)+ \frac{1}{4!}\zeta_{0}\,{g_{\omega}^2}\,\omega^4 
\nonumber\\
&&
+\frac{1}{2}m_{\omega}^2\omega^2\Bigg(1+\eta_{1}\frac{g_\sigma\sigma}{M_{\rm nucl.}}+\frac{\eta_{2}}{2}\frac{g_\sigma^2\sigma^2}{M_{\rm nucl.}^2}\Bigg)
\nonumber\\
&&
+ \frac{1}{2}\Bigg(1+\frac{\eta_{\rho}g_\sigma\sigma}{M_{\rm nucl.}}\Bigg)m_{\rho}^2\,\rho^{2}-\frac{1}{2}m_{\delta}^2\, \delta^{2}+\Lambda_{\omega} g_\rho^2 g_\omega^2 \rho^2 \omega^2
\nonumber\\
&&
+\sum_{j=e,\mu}  \frac{g_s}{3(2\pi)^{3}}\int_{0}^{k_{f_{j}}} \frac{k^2}{\sqrt{k^2 + m^2_{j}}} \, d^{3}k.
\end{eqnarray}
where mass of the nucleon is $M$ and $g_s$ is the spin degeneracy. The $m_\sigma$, $m_\omega$, $m_\rho$, and $m_\delta$ are the masses $\sigma$, $\omega$, $\rho$, and $\delta$ mesons respectively. The corresponding coupling constants are represented as  $g_\sigma$, $g_\omega$, $g_\rho$, and $g_\delta$. The self-interactions and cross-coupling terms among mesons are represented as  $\kappa_3$, $\kappa_4$, $\zeta_0$ and  $\eta_1$, $\eta_2$, $\eta_\rho$,  $\Lambda_\omega$ respectively \cite{ERMF6, Serot1992, ERMF7, E-RMF8, E-RMF9, E-RMF5, PhysRevC.97.045806}.

\subsection{Quarkyonic Model}
\label{b}
The quarkyonic model where it is assumes the matter density is several times higher than the nuclear saturation density at the NS core. \cite{PhysRevLett.122.122701} At such high densities nucleons break into quarks near some transition density. This model's characteristic feature is the sound peak's appearance near the transition density. The emergence of the quarkyonic phase from nuclear matter is signalled by a rapid increase in pressure, attributed to quarks filling lower momentum levels once the baryon density reaches a certain critical transition density. This change allows low-momentum states to be regarded as quark-based, while high-momentum states near the Fermi surface continue to behave as nucleonic. Momentum states close to the Fermi surface are comparable to the QCD confinement scale, $\Lambda_{\rm cs}$, forming nucleons as quark-bound states. \\
The foundational model proposed by McLerran and Reddy \cite{PhysRevLett.122.122701} was further extended and refined by Zhao and Lattimer \cite{PhysRevD.102.023021}. They took into account the conditions for beta-equilibrium and charge neutrality within quarkyonic matter. Nucleons interact through a potential energy dependent on nucleon density, calibrated to match specific properties of uniform nuclear matter. Additionally, they achieved chemical equilibrium among neutrons, protons, and quarks, establishing a link between $k_{f_{n,p}}$ and $k_{f_{u,d}}$, highlighting the unique characteristics of this modified quarkyonic model. In this model, nucleons occupy a Fermi shell, giving rise to a defined minimum Fermi momentum $k_{f0_{(n,p)}}$ and an upper Fermi momentum $k_{f_{n,p}}$. The Fermi momenta for the $d$ and $u$ quarks are $k_{f_d}$ and $k_{f_u}$, respectively.


The conservation law of baryon density gives us,\cite{PhysRevD.102.023021},
\begin{eqnarray}
n &=& n_n + n_p + \frac{n_u+n_d}{3}
\nonumber\\
&=& \frac{g_s}{6\pi^2}\bigg[(k_{f_{n}}^3-k_{0_{n}}^3)+(k_{f_{p}}^3-k_{0_{p}}^3)+\frac{(k_{f_{u}}^3+k_{f_{d}}^3)}{3}\bigg],  
\end{eqnarray}
also, nucleons, quarks, and leptons respect the charge neutrality condition and that gives us,
\begin{eqnarray}
n_p + \frac{2n_{u}}{3} -  \frac{n_{d}}{3}& = & n_{e^{-}} + n_{\mu}.  
\end{eqnarray}
The minimum momentum for protons and neutrons is connected to the transition Fermi momentum $k_{\rm Ft}$, which corresponds to the transition density $n_t$, and is related to their respective Fermi momenta as given by the following expression \cite{PhysRevD.102.023021}.
\begin{eqnarray}
 k_{0(n,p)} &=& (k_{f_{(n,p)}}-k_{t_{(n,p)}})\bigg[1+ \frac{\Lambda_{cs}^2}{k_{f_{(n,p)}}k_{t_{(n,p)}}}\bigg].
\end{eqnarray}
\noindent
The equilibrium of the strong interaction guarantees that the Fermi gas attains its minimum possible energy at a specified baryon density. This is equivalent to achieving chemical equilibrium between nucleons and quarks, represented as follows \cite{PhysRevD.102.023021}.
\begin{eqnarray}\label{ebnq}
\mu_n &=& \mu_u + 2\mu_d, \\
\mu_p &=& 2\mu_u + \mu_d.
\end{eqnarray}
\noindent
Here the chemical potentials of the neutron, proton, up quark, and down quark are $\mu_n$, $\mu_p$, $\mu_u$, and $\mu_d$ respectively. The energy of the Fermi gas is further reduced by meeting the beta-equilibrium condition while ensuring charge neutrality. This establishes chemical equilibrium among neutrons, protons, electrons, and muons, as represented by \cite{Glendenning,PhysRevD.102.023021}.
\begin{eqnarray} \label{qnbe}
\mu_{n} &=& \mu_{p} + \mu_{e^{-}}, \nonumber \\
\mu_{\mu} &=& \mu_{e^{-}}.
\end{eqnarray}
\noindent
A key feature of this model is that the masses of the up and down quarks are not independent; rather, they are influenced by the beta-equilibrium condition in neutron star matter. These quark masses arise at the transition density $n_t$ and are determined by the chemical potentials of neutrons and protons, $\mu_{t_n}$ and $\mu_{t_p}$, respectively. The expressions for the quark masses are given as follows.
\begin{eqnarray}
m_u &=& \frac{2}{3} \mu_{t_p} - \frac{1}{3} \mu_{t_n}, \nonumber \\
m_d &=& \frac{2}{3} \mu_{t_n} - \frac{1}{3} \mu_{t_p},
\end{eqnarray}
\noindent
Here, $\mu_{t_n}$ and $\mu_{t_p}$ denote the chemical potentials of neutrons and protons at the transition density $n_t$, within beta-equilibrium matter comprising interacting neutrons, protons, electrons, and muons in a mesonic mean-field framework.
\noindent
Since in this model quarks are non-interacting fermion gas, quark energy density energy density ${\cal E}_{\rm QM}$ and pressure $P_{\rm QM}$ can be expressed as
\cite{PhysRevD.102.023021}.
\begin{eqnarray}
{\cal E}_{\rm QM}&=& \sum_{j=u,d}\frac{g_s N_c}{(2\pi)^3}\int_0^{k_{f_{j}}}k^2\sqrt{k^2 + m_{j}^2 }\, d^3k,
\end{eqnarray}
\begin{eqnarray}
P_{\rm QM} &=& \mu_{u} n_{u} + \mu_{d} n_{d} - \epsilon_{QM} \, .
\end{eqnarray}

\subsection{Dark Matter Model} \label{c}
This work examines a basic Dark Matter (DM) model, where DM particles engage with nucleons and quarks through Higgs exchange. The Lagrangian density for this interaction is outlined in \cite{DM1,DM2,DM3},
\begin{eqnarray}
{\cal{L}}_{\rm DM} &=& \bar{\chi} \left[ i \gamma^\mu \partial_\mu - M_\chi + y h \right] \chi 
+ \frac{1}{2} \partial_\mu h \partial^\mu h \nonumber \\
&& - \frac{1}{2} M_h^2 h^2 + f \frac{M_{\text{nucl.}/u/d}}{v} \bar{\psi} h \psi.
\label{eq:LDM}
\end{eqnarray}
\noindent
Here, the wave functions for the DM particle and nucleons are represented by $\chi$ and $\psi$, respectively. The interaction between the Higgs boson and nucleons follows a Yukawa-like form, with $f$ being the coupling constant, which corresponds to the proton-Higgs form factor. For the DM particle, the Neutralino is chosen, with a mass $M_\chi$ set at 200 GeV. The values of $y$ and $f$ are selected as 0.07 and 0.35, respectively, based on various experimental and empirical constraints \cite{DM5}. The Higgs boson mass ($M_h$) is assumed to be 125 GeV, and its vacuum expectation value ($v$) is 246 GeV.

The energy density and pressure for the DM after relativistic mean-field approximation can be given as  \cite{DM1, DM3, DM2, mnras}
\begin{eqnarray}
{\cal{E}}_{\rm DM}& = & \frac{2}{(2\pi)^{3}}\int_0^{k_f^{\rm DM}} d^{3}k \sqrt{k^2 + (M_\chi^\star)^2} + \frac{1}{2}M_h^2 h_0^2 ,
\label{eq:EDM}
\end{eqnarray}
\noindent
and
\begin{eqnarray}
P_{\rm DM}& = &\frac{2}{3(2\pi)^{3}}\int_0^{k_f^{\rm DM}} \frac{d^{3}k k^2} {\sqrt{k^2 + (M_\chi^\star)^2}} - \frac{1}{2}M_h^2 h_0^2 ,
\label{eq:PDM}
\end{eqnarray}
\noindent
where $k_f^{\rm DM}$ is defined to be DM Fermi momentum. Assuming that the nucleon density is 1000 times larger than the average DM density, the resulting mass ratio is $M_{\chi}/M_{\rm NS} = 1/6$. Based on this assumption, the Fermi momentum for DM can be approximated as $k_f^{\rm DM} \approx 0.03$ GeV. The effective masses of the nucleons, which are altered by their interaction with the Higgs field, along with the effective mass of DM, are given by the following expressions:
\begin{eqnarray}
M_i^\star &=& M_{\rm nucl.} + g_\sigma \sigma_0 \mp g_\delta \delta_0 - \frac{f M_{\rm nucl./u/d}}{v} h_0, 
\nonumber\\
M_\chi^\star &=& M_\chi - y h_0,
\label{eq:effm_tot}
\end{eqnarray}
\noindent
Here, $\sigma_0$, $\delta_0$, and $h_0$ represent the mean-field values of the mesons and the Higgs field, respectively. The total energy and pressure of a quarkyonic star containing DM are then expressed as follows:
\begin{eqnarray}
\cal{E} &=& {\cal{E}}_{\rm BM} + {\cal{E}}_{\rm QM} + {\cal{E}}_{\rm DM},
\label{eq:effm_total}
\end{eqnarray}
\begin{eqnarray}
P &=& P_{\rm BM} + P_{\rm QM} + P_{\rm DM},
\label{eq:press_total}
\end{eqnarray}
Here, ${\cal{E}}_{\rm BM}$, ${\cal{E}}_{\rm QM}$, and ${\cal{E}}_{\rm DM}$ represent the energy densities of nucleonic matter, quark matter, and dark matter, respectively, and $P_{\rm BM}$, $P_{\rm QM}$, and $P_{\rm DM}$ are their corresponding pressures. The total energy and pressure of a DM-admixed quarkyonic star are then given by the combination of these contributions.


\begin{figure*}
\centering
\includegraphics[width=1.5\columnwidth]{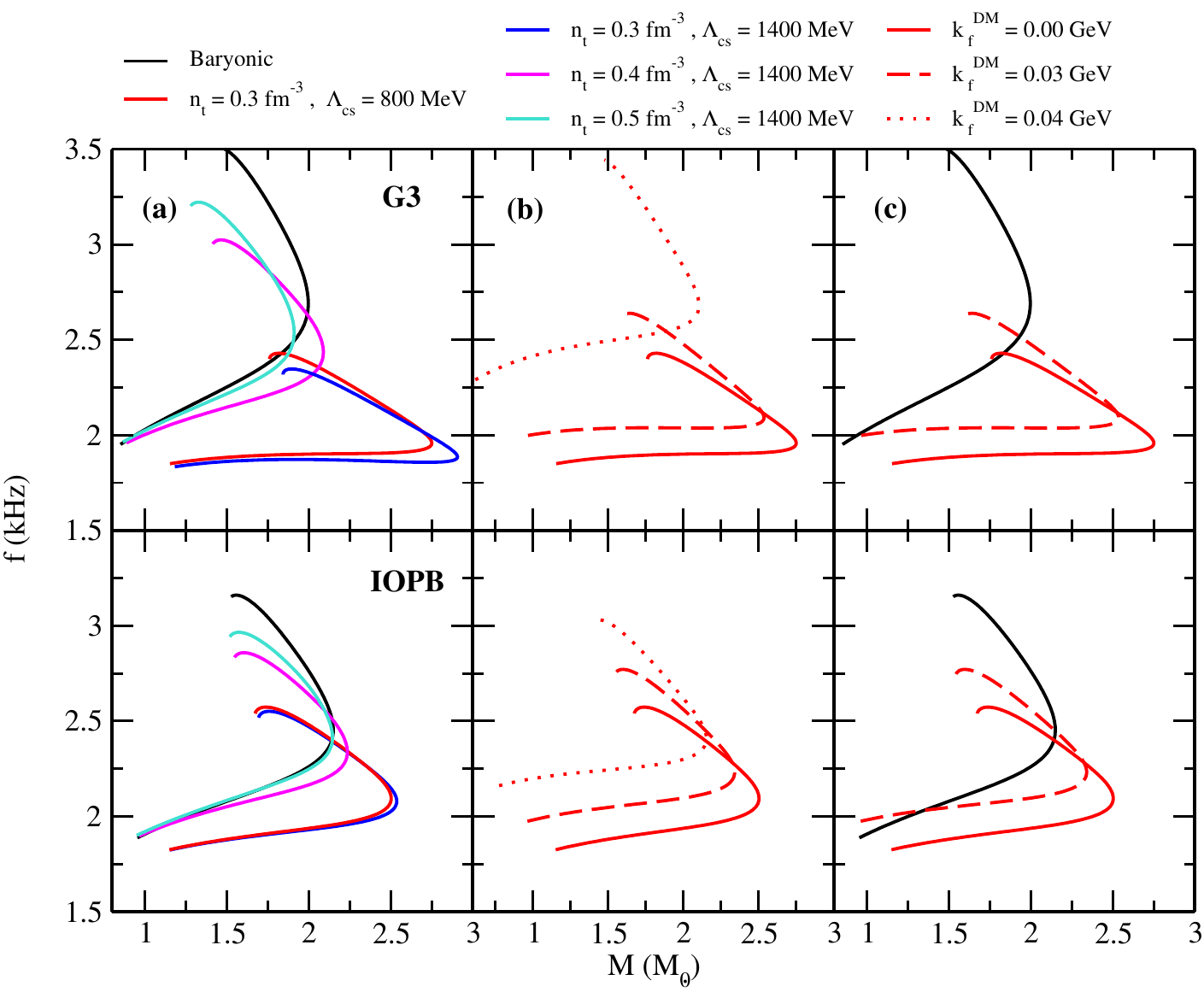}
\caption{$f-M$ relations for (a) baryonic and QM (b) DM admixed with QM (c) three individual cases of baryonic, QM and DM admixed QM for NM parameter G3 and IOPB-I. } 
\label{fig1}
\end{figure*}

\subsection{The equilibrium state} \label{d}
The metric for the non-rotating and spherically symmetric space-time is expressed as 
\begin{equation}
ds^2 = -e^{\nu} dt^2 + e^{\lambda} dr^2 + r^2 \left( d\theta^2 + \sin^2\theta \, d\phi^2 \right),
\end{equation}

where $\nu$ and $\lambda$ are functions of $r$. The equilibrium configuration of non-rotating relativistic stars are spherically symmetric solutions of the well-known Tolman-Oppenheimer-Volkoff (TOV) equations \cite{TOV1, TOV2} given as:
\begin{equation}
  \frac{dm}{dr} = 4\pi r^2\cal{E},
\end{equation}

where $\cal{E}$ is the energy density and m is the enclosed mass, while the TOV equations which determine pressure $P(r)$ and metric function $\Phi(r)$ are described as
\begin{align}
   \frac{dP}{dr} & =-( \mathcal{E}+ P) \frac{d\nu}{dr} \\
  \frac{d\nu}{dr} &= 2 \frac{M+4\pi r^3 P}{r(r-2M)}.
\end{align}
An additional equation is needed to close the system of equations, that is, the EOS computed from microscopic consideration.

\subsection{Perturbation equation} \label{e}
In this section, we derive the perturbation equations for the non-radial oscillations of neutron stars within the Cowling approximation \cite{Cowling1, cowling01, cowling02}. In the Cowling approximation, the space-time metric is assumed to remain fixed. Despite this simplification, the Cowling formalism provides a sufficiently accurate description, yielding the oscillation spectrum with good precision. Comparisons between oscillation frequencies obtained using a fully general relativistic numerical approach and those derived from the Cowling approximation shows that the error is less than $20\%$. The perturbed metric is expressed as follows:
\begin{equation}
\begin{split}
ds^2 = - [1+r^l H_0(r)e^{i\omega t} Y_{lm}(\phi,\theta)] e^{\nu(r)} dt^2\\
+ [1-r^l H_0(r)e^{i\omega t} Y_{lm}(\phi,\theta)] e^{\lambda(r)}dr^2 \\
+ [1-r^l K(r)e^{i\omega t}Y_{lm}(\phi,\theta)]r^2 d\Omega^2\\
-2i\omega r^{l+1}H_1(r)e^{i\omega t}Y_{lm}(\phi,\theta) dt~dr\,,
\end{split}
\end{equation}
where 
\begin{equation}
e^{\lambda(r)}=\frac{1}{1-\frac{2  m(r)}{r}}
\end{equation}
and 
\begin{align}
e^{\nu(r)} = \exp \bigg(-2 \int_{0}^{r}\left\{\frac{\left[m(r^{\prime})+4 \pi p(r^{\prime}) r^{\prime 3}\right]}{r^{\prime}\left[r^{\prime}-2 m(r^{\prime}) \right]}\right\} d r^{\prime}\bigg){\rm e}^{\nu_0},
\label{eq:static_metric_nu}
\end{align}
Here, $m(r^\prime)$ represents the mass of the star enclosed at radius $r^\prime$, and $p$ is the corresponding pressure. The radial and angular perturbation of the metric is given by functions $H_0$, $H_1$, $K$ and spherical harmonics $Y_{lm}$, where $l$ and $m$ are the orbital angular momentum number and the azimuthal number respectively. The complex oscillation frequency is defined by the quantity $\omega$ whose real part represents the f-mode oscillation frequency and the imaginary part is associated with the inverse of the damping time. The Lagrangian displacement vector for the perturbed fluid element is given by  
\begin{eqnarray}
\xi^r &=& r^{l-1}e^{-\frac{\lambda}{2}}W Y^l_m e^{i\omega t} \label{eq:xi_radial}\\
\xi^\theta &=& -r^{l-2} V \partial_\theta Y_m^l e^{i\omega t}\\
\xi^\phi &=& -\frac{r^{l-2}}{ \sin^{2}\theta} V\partial_\phi Y_m^l e^{i\omega t} \,,
\end{eqnarray}
The perturbation amplitude W and V are defined by the above equations. The  oscillation equation within relativistic cowling approximation is derived by setting $H_0 = H_1 = K = 0$. These assumptions result in:
\begin{eqnarray}
\label{eq:ODE_DL3_cowling_UW}
\frac{dW}{d\ln r} && = -(l+1)\left[W-l e^{\nu +\lambda/2} U\right] \nonumber \\
&& -\frac{e^{\lambda/2}(\omega r)^2}{c_{ad}^2}\left[U-\frac{d\Phi}{d \ln r} \frac{e^{-\lambda/2}}{(\omega r)^2} W\right ]\,,\\
\frac{dU}{d\ln r} && = e^{\lambda/2-\nu}\left[W -le^{\nu-\lambda/2}U\right] \nonumber \\
&& + (\frac{1}{c_{\rm eq}^2}-\frac{1}{c_{\rm ad}^2} )\frac{d\Phi}{d \ln r} \left [ U -\frac{d\Phi}{d \ln r} \frac{e^{-\lambda/2}}{(\omega r)^2} W\right ]\,.\label{eq:ODE_DL4_cowling_UW}
\end{eqnarray}
 Here, $W = e^{\lambda/2} r^{1-l} \xi^r$, $U = -e^{-\nu}V$ and $\Phi = 2\nu$.
The boundary conditions can be written explicitly as,
\begin{eqnarray}
\left.\frac{W}{U}\right|_{r=0}&=&l e^{\nu|_{r=0}} \\
\left.\frac{W}{U}\right|_{p=0}&=&\frac{\omega^2R^3}{M}\sqrt{1-\frac{2GM}{R}} \,.
\end{eqnarray}
Solving these equations gives us the f-mode oscillation frequency of the Neutron star within the relativistic cowling approximation.

\begin{figure}
\centering
\includegraphics[width=1\columnwidth]{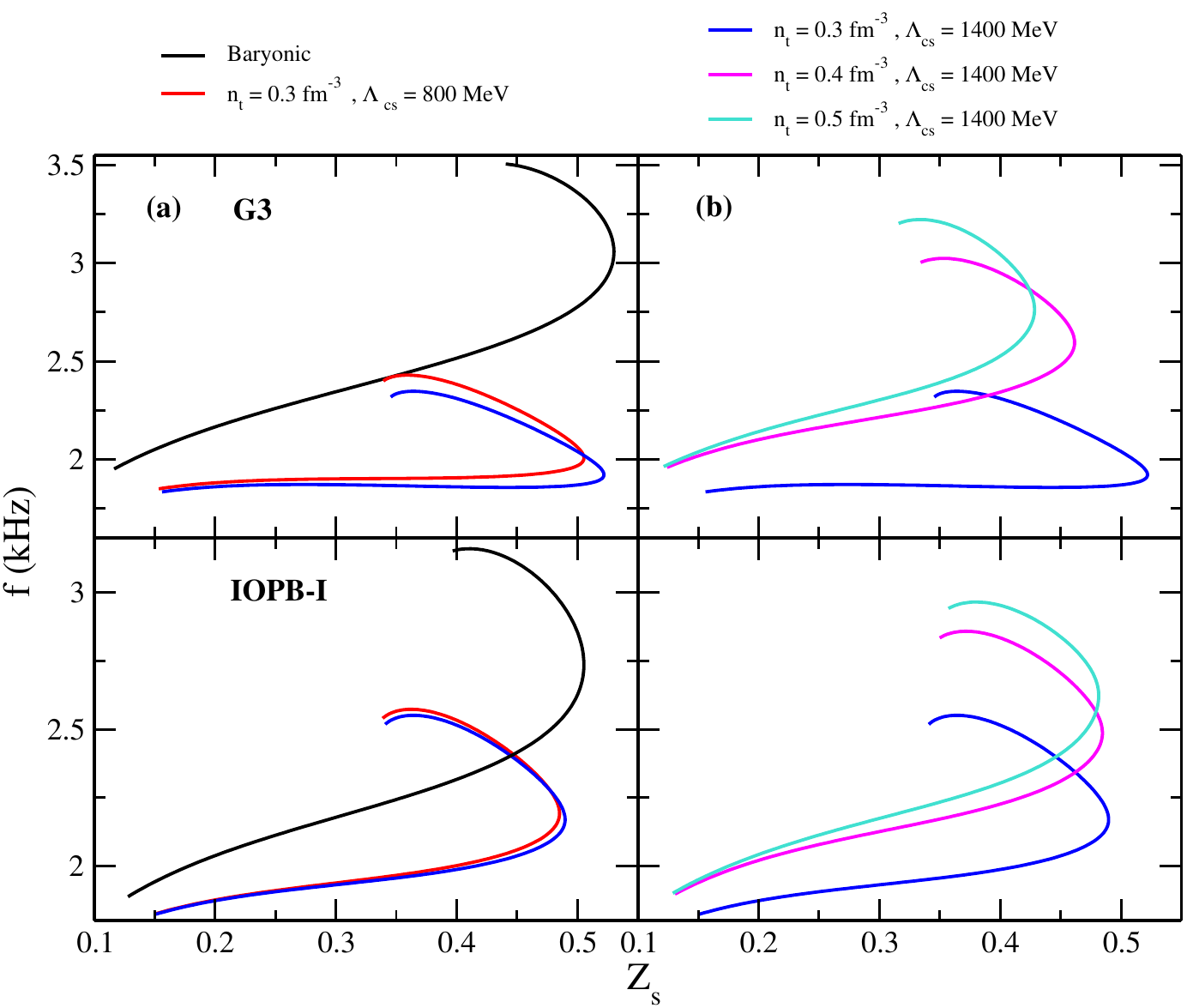}
\caption{$f-Z_s$ relations for (a) baryonic and QM at fixed transition density $n_t = 0.3 $ $\rm fm^{-3}$ (b) QM at fixed confinement scale $\Lambda_{cs} = 1400 $ MeV for parameter G3 and IOPB-I respectively.}
\label{fig2}
\end{figure}
\section{Results and Discussions}
\label{results}
In this section, we calculate the f-mode oscillation frequencies for quadrupole deformations (l = 2) as functions of various stellar parameters, including mass (M), compactness (C), redshift ($Z_{s}$), the star's average density $(\bar{M}/\bar{R}^3)^{1/2}$, and tidal deformability ($\Lambda$). These calculations are performed using RMF EOSs for dark matter-admixed quarkyonic stars. The EOSs for baryonic matter, quarkyonic matter, and dark matter-admixed quarkyonic matter under beta equilibrium are described. In Table \ref{tab1} we have listed all the macroscopic properties along with f-mode frequency for all the EOSs. Furthermore, the universal relations related to $f-$mode frequency along with correlation among various macroscopic properties of NS is studied.

\begin{table*}
\centering
\renewcommand{\tabcolsep}{0.05cm}
\renewcommand{\arraystretch}{1.5}
\caption{DM admixed quarkyonic star properties for G3 and IOPB-I parameter sets \cite{G3, IOPB-I}. }
\begin{tabular}{cccccccc     cc cc cc cc}
\hline \hline
Model  & $n_{\rm t}$ & $\Lambda_{\rm cs }$ & $k_f^{\rm DM}$ & $M_{\rm max}$ & $R_{1.4}$   & $\Lambda_{max}$     & $f_{max}$ & $f_{1.4}$   \\
       & (fm$^{-3}$) & (MeV)               & (GeV)          & ($M_\odot$)   & (km)        &      & (kHz)             & (kHz)       &           &            &           &   \\
\hline
G3 & 0.00 & 0.00  & 0.00 &  1.997& 12.11 & 464.63    & 2.687   & 2.214      \\
G3 & 0.3 & 800  & 0.00 & 2.75 & 14.17 & 1181.62      & 1.887   & 1.857       \\
G3 & 0.3 & 800  & 0.03 & 2.54 & 13.24 & 838.30       & 2.348   & 2.207      \\
G3 & 0.3 & 800  & 0.04 & 2.10 & 11.26 & 290.19       & 2.675   & 2.443        \\[0.2cm] 
G3 & 0.3 & 1400 & 0.00 & 2.91 & 14.29 & 1251.80      & 1.957   & 1.877          \\
G3 & 0.3 & 1400 & 0.03 & 2.14 & 12.44 & 535.60       & 2.090   & 2.034             \\
G3 & 0.3 & 1400 & 0.04 & 1.85 & 11.41  & 317.15      & 2.680   & 2.483        \\[0.2cm] 
G3 & 0.4 & 1400 & 0.00 & 2.09  & 12.9   & 599.05     & 2.432   & 2.124           \\
G3 & 0.4 & 1400 & 0.03 & 1.99 & 12.08  & 424.00      & 2.567   & 2.289          \\
G3 & 0.4 & 1400 & 0.04 & 1.90   & 11.27  & 291.60    & 2.720   & 2.481       \\[0.2cm] 
G3 & 0.5 & 1400 & 0.00 & 1.91 & 12.64 & 505.60       & 2.534   & 2.182                    \\
G3 & 0.5 & 1400 & 0.03 & 1.82 & 11.79 & 352.57       & 2.675   & 2.362             \\
G3 & 0.5 & 1400 & 0.04 & 1.78 & 10.97  & 232.43      & 2.851   & 2.567           \\
\hline \hline
IOPB-I & 0.00 & 0.00  & 0.00 & 2.149 & 12.78 & 689.62      &  2.456     & 2.054     &           &      &           &        &\\
IOPB-I & 0.3 & 800  & 0.00 & 2.50 & 14.26 & 1159.70        &  2.080     & 1.865     &           &      &           &        & \\
IOPB-I & 0.3 & 800  & 0.03 & 2.34 & 13.26 & 796.46         &  2.218     & 2.034     &           &      &           &        &   \\
IOPB-I & 0.3 & 800  & 0.04 & 2.15 & 12.25 & 519.72         &  2.385     & 2.231     &           &      &           &        & \\[0.2cm] 
IOPB-I & 0.3 & 1400 & 0.00 & 2.54 & 14.27  & 1169.31       &  2.094     & 1.870     &           &      &           &        &\\
IOPB-I & 0.3 & 1400 & 0.03 & 2.37 & 13.28  & 804.95        &  2.232     & 2.039     &           &      &           &        &       \\
IOPB-I & 0.3 & 1400 & 0.04 & 2.19  & 12.28   & 530.93      &  2.406     & 2.237     &           &      &           &        &\\[0.2cm] 
IOPB-I & 0.4 & 1400 & 0.00 & 2.24 & 13.4   & 743.63        &  2.337     & 2.030     &           &      &           &        & \\
IOPB-I & 0.4 & 1400 & 0.03 & 2.12  & 12.49  & 515.81       &  2.471     & 2.205     &           &      &           &        &  \\
IOPB-I & 0.4 & 1400 & 0.04 & 1.98 & 11.58  & 342.53        &  2.637     & 2.405     &           &      &           &        &    \\[0.2cm] 
IOPB-I & 0.5 & 1400 & 0.00 & 2.15 & 13.27 &  685.78        &  2.418     & 2.060     &           &      &           &        &    \\
IOPB-I & 0.5 & 1400 & 0.03 & 2.04  & 12.34  & 467.08       &  2.554     & 2.240     &           &      &           &        &     \\
IOPB-I & 0.5 & 1400 & 0.04 & 1.92 & 11.42 & 304.12         &  2.720     & 2.448     &           &      &           &        &       \\
\hline \hline
\end{tabular}
\label{tab1}
\end{table*}

\subsection{f-mode oscillation frequencies as a function of stellar parameters}
\label{ra}

\begin{figure}
\centering
\includegraphics[width=1\columnwidth]{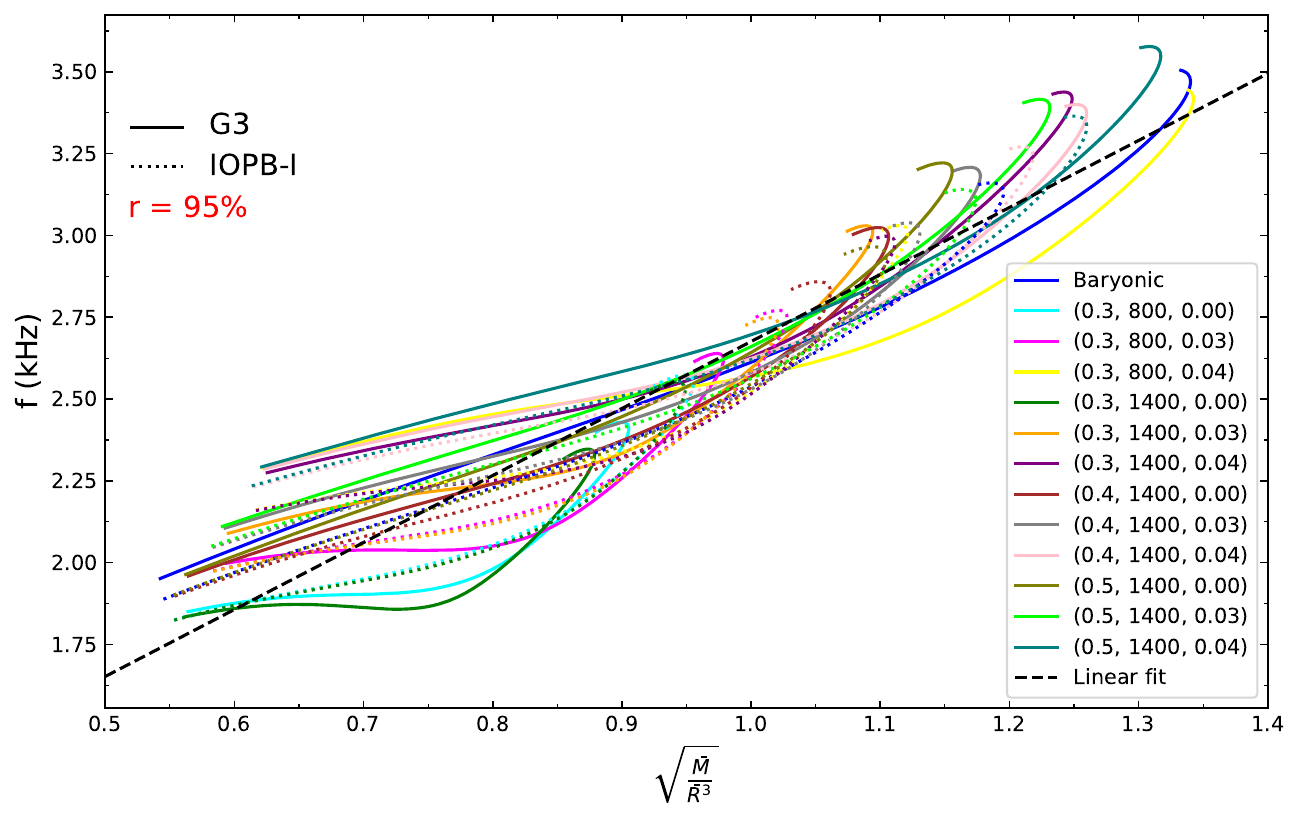}
\caption{ f-mode frequency as a function of average density $(\bar M/\bar R^3)^\frac{1}{2}$ for varying ($\rm n_t$, $\Lambda_{\rm cs}$, $\rm k_f^{DM}$) ($\rm fm^{-3}$, $\rm MeV$, $\rm GeV$) for G3 and IOPB-I parameters.}
\label{fig3}
\end{figure}
\begin{figure}
\centering
\includegraphics[width=1\columnwidth]{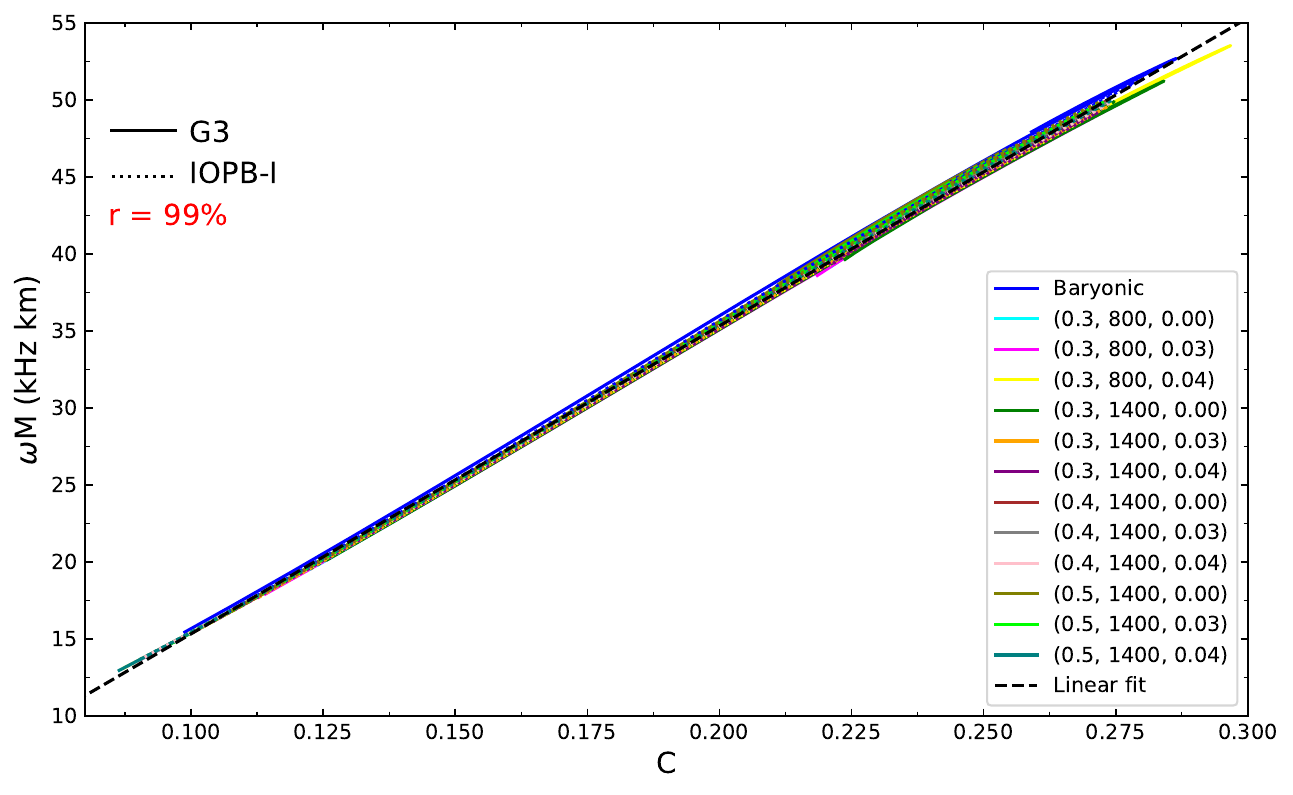}
\caption{Scaled f-mode frequency $\omega M$ as a function of steller compactness $C = M/R$ for varying ($\rm n_t$, $\Lambda_{\rm cs}$, $\rm k_f^{DM}$) ($\rm fm^{-3}$, $\rm MeV$, $\rm GeV$) for G3 and IOPB-I parameters.
}\label{fig4}
\end{figure}
\begin{figure}
\centering
\includegraphics[width=1\columnwidth]{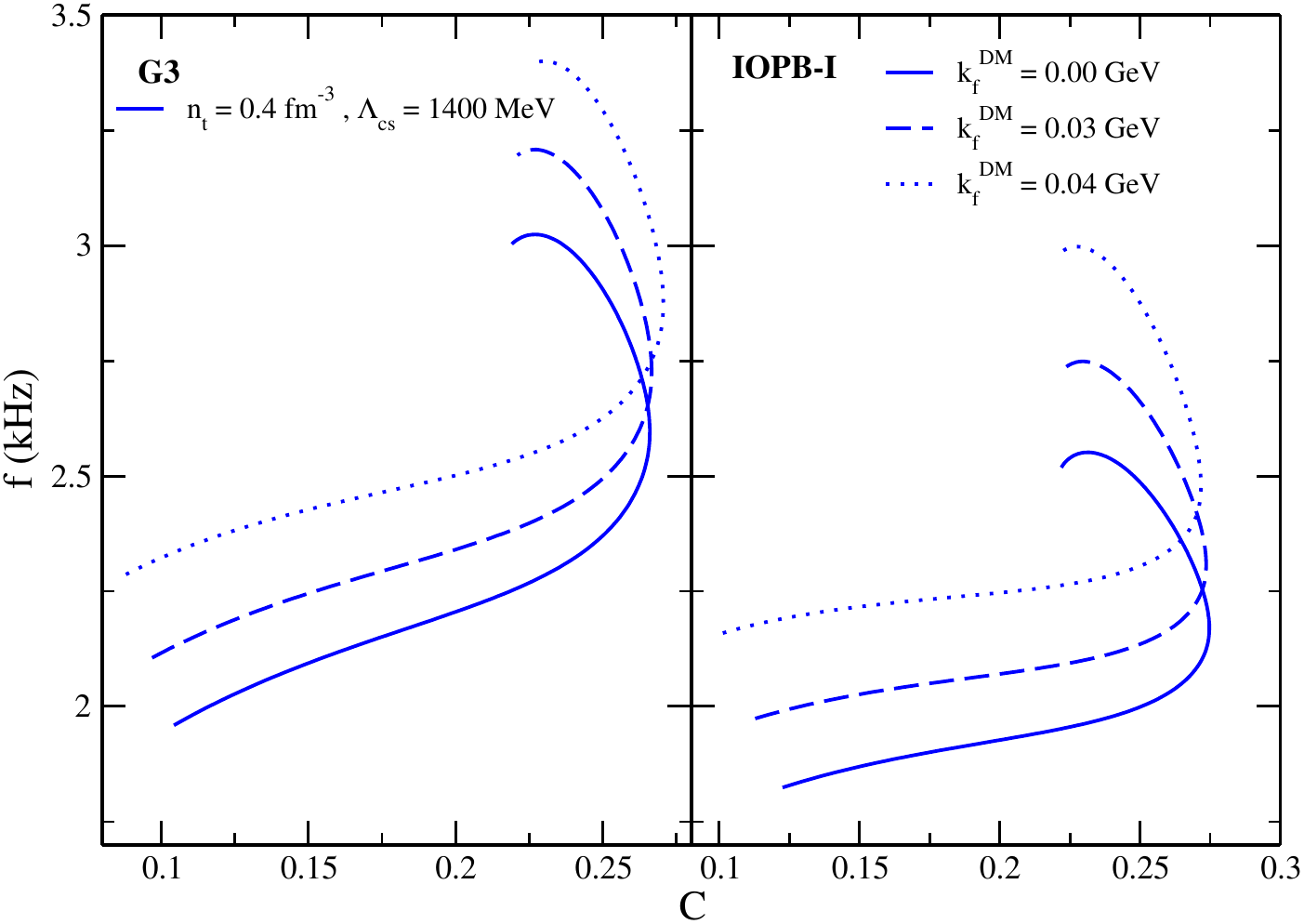}
\caption{f mode frequency as a function of C at fixed $n_t = 0.4$  $\rm 
 fm^{-3}$, $\Lambda_{cs} = 1400 $ MeV for (a) G3 and (b) IOPB-I at varying dark matter percentage.}
\label{fig5}
\end{figure}

\begin{figure*}
\label{C_plot}
\centering
\includegraphics[width=1\columnwidth]{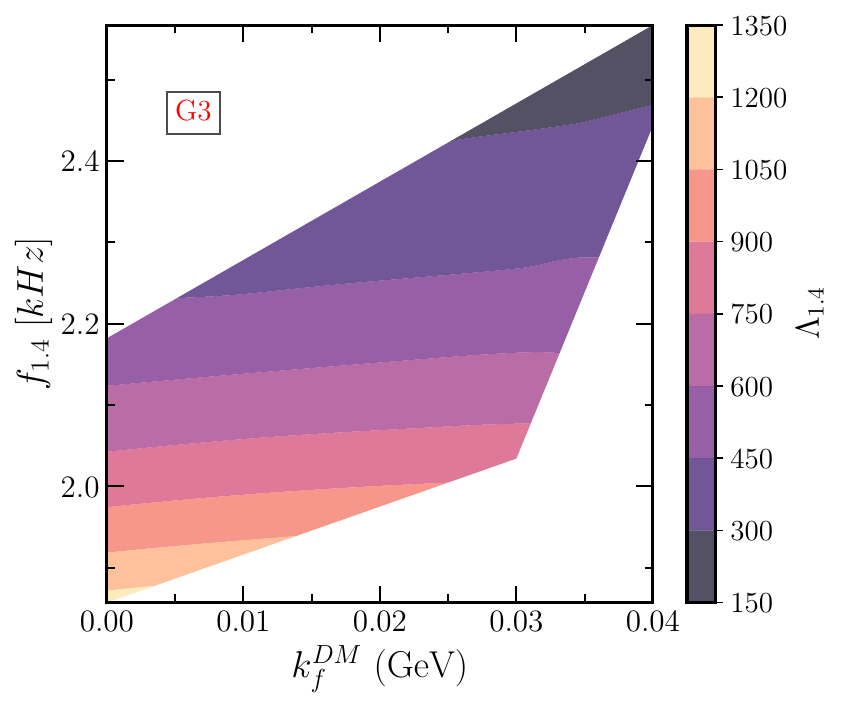}
\includegraphics[width=1\columnwidth]{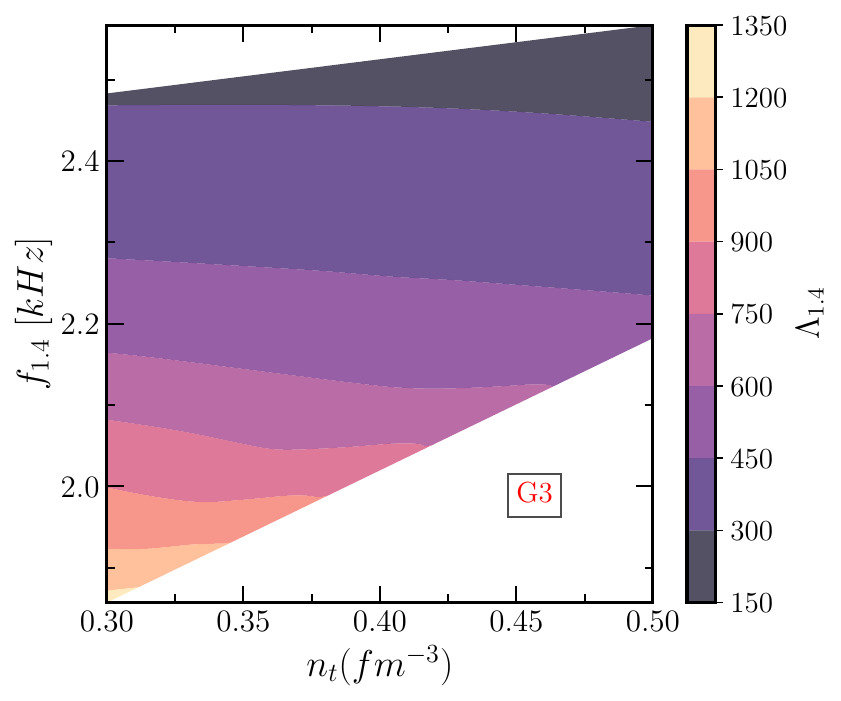}
\includegraphics[width=1\columnwidth]{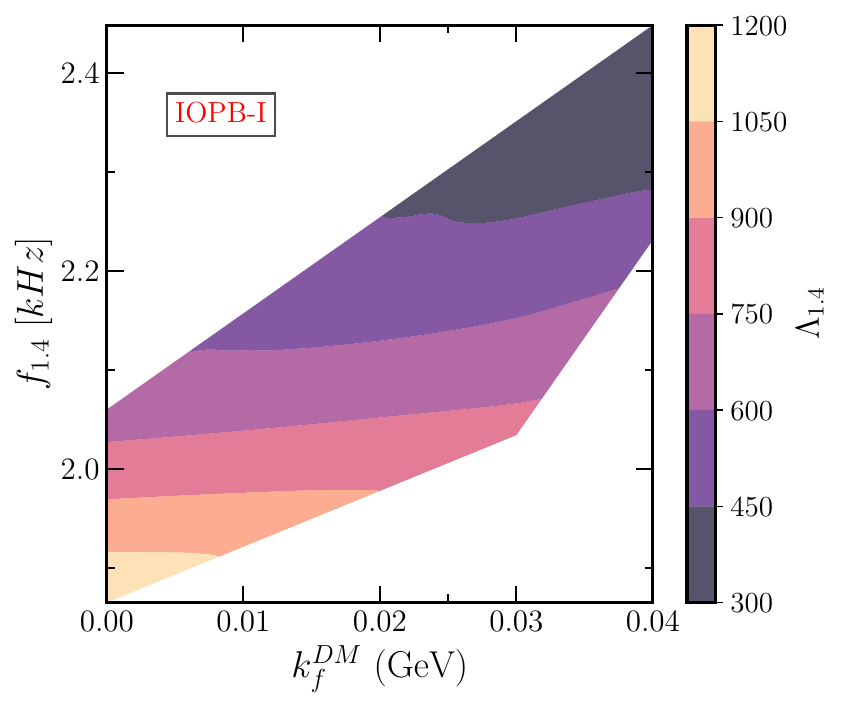}
\includegraphics[width=1\columnwidth]{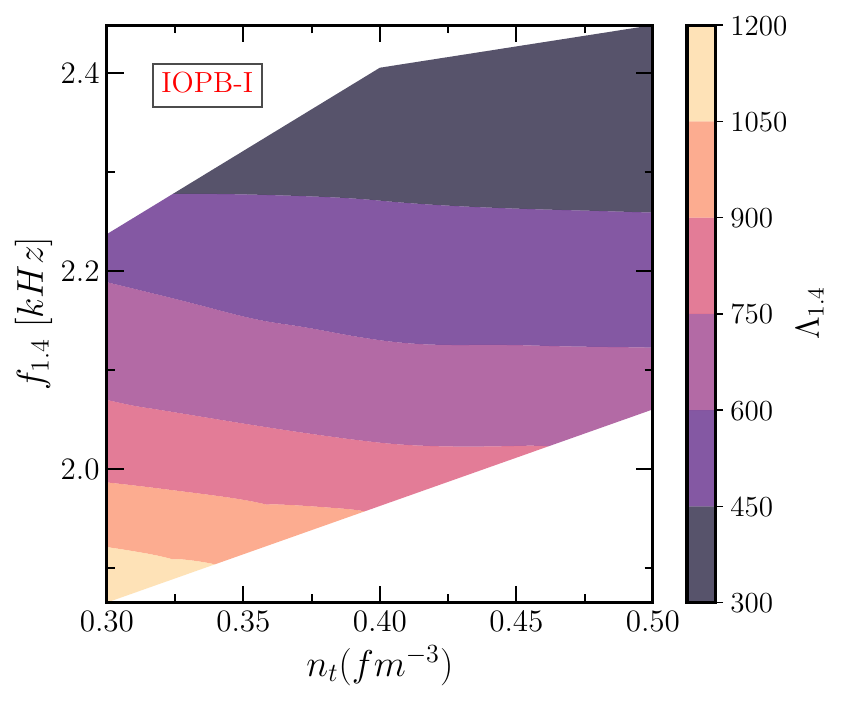}
\caption{$f_{1.4}$ as a function of DM momenta $k_f^{DM}$ for G3 and IOPB-I NM parameter sets for a specific quarkyonic star ($n_t =0.3$  $\rm fm^{-3}$, $\Lambda_{cs} = 1400$ MeV  ) (left panel) while in the right panel $f_{1.4}$ as a function of $n_t$ is shown for a fixed  $k_f^{DM} = 0.03 $ GeV. The color bar represents the tidal deformability ($\Lambda_{1.4}$). }
\label{fig6}
\end{figure*}
\begin{figure}
\centering
\includegraphics[width=1\columnwidth]{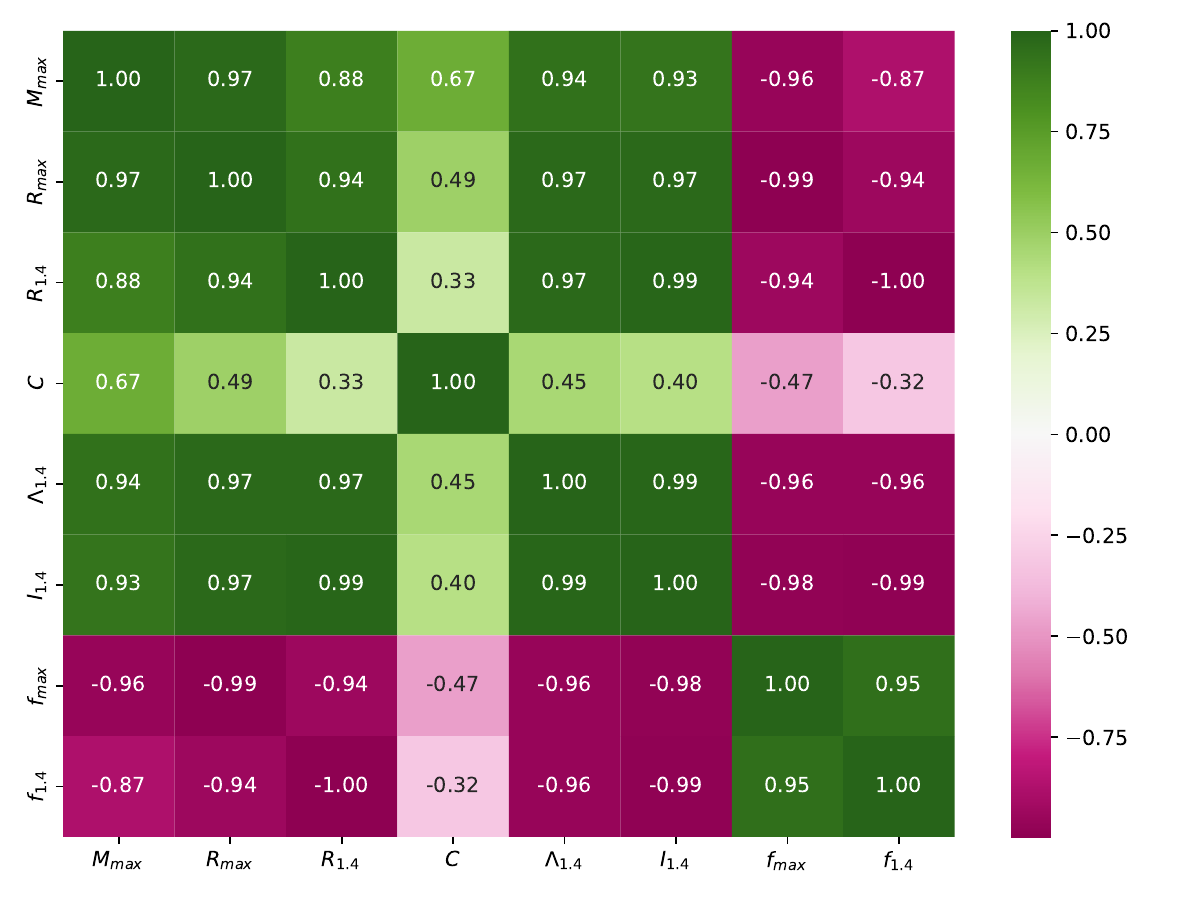}
\caption{Correlation heat map of the various macroscopic properties of NS ( $n_t =0.3$ ,$\rm fm^{-3}$, $ \Lambda_{cs} =1400 $ MeV , $k_f^{DM} = 0.03 $ GeV). The color map indicates the intensity of the correlation, while the numbers show the corresponding p-values.  }
\label{fig7}
\end{figure}
\begin{figure}
\centering
\includegraphics[width=1\columnwidth]{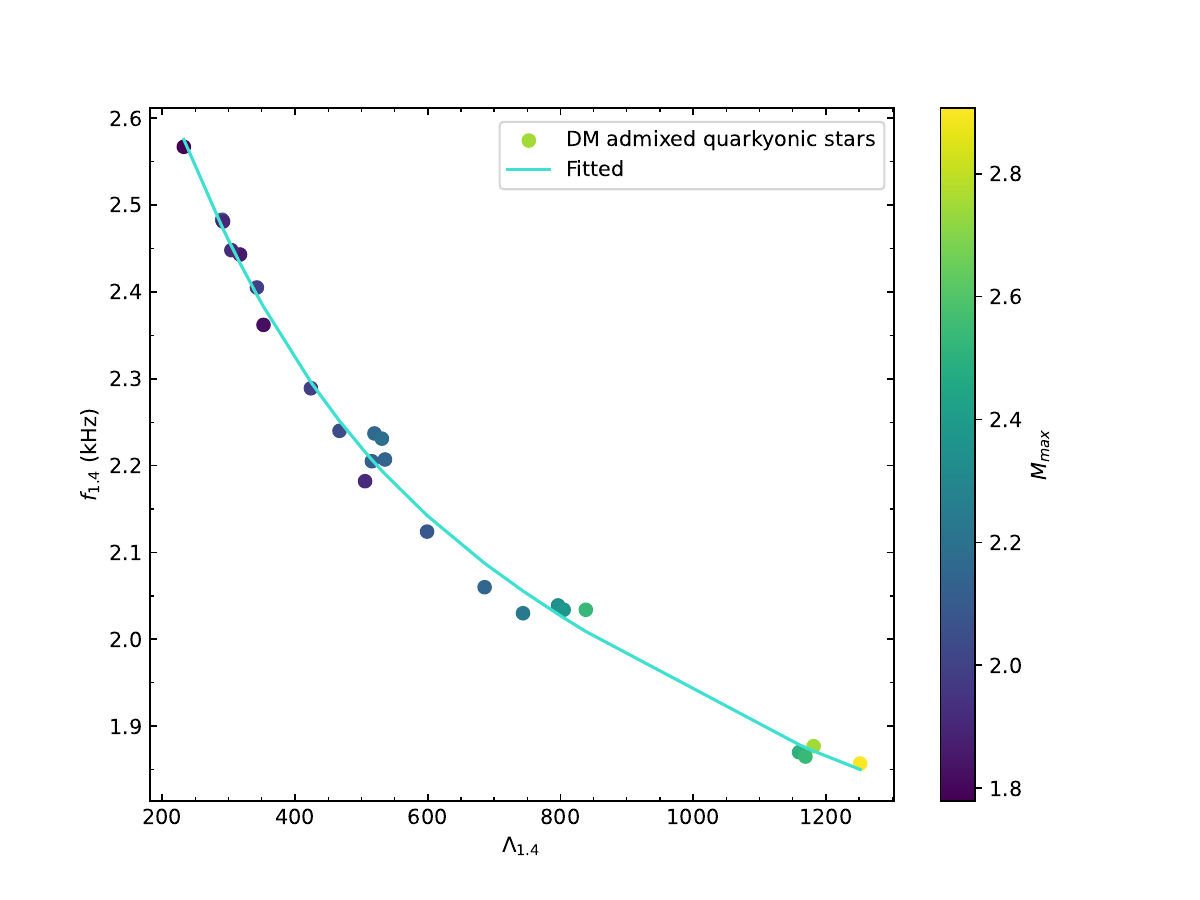}
\includegraphics[width=1\columnwidth]{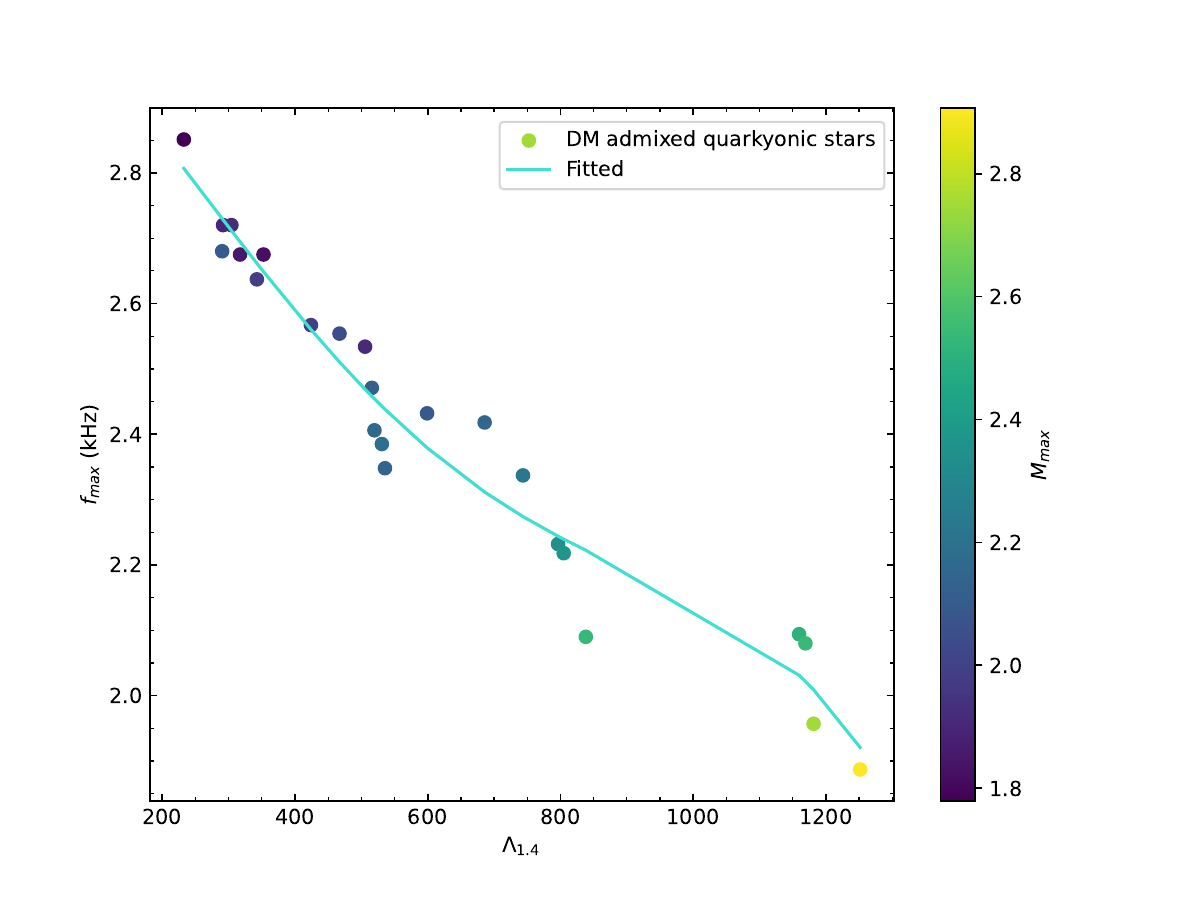}
\caption{Left: Correlation between $f_{1.4}$-$\Lambda_{1.4}$ (top) and IOPB-I $f_{max}$-$\Lambda_{1.4}$  for all the combination of free parameter $n_t$, $\Lambda$, $k_f^{DM}$ for IOPB-I. The color bar
represents the maximum masses of the corresponding parameter sets.}
\label{fig8}
\end{figure}

\begin{itemize}
\item In Fig. \ref{fig1}, we present the f-mode frequency versus mass plot. Panel (a) compares baryonic matter and quarkyonic matter with beta equilibrium. We observe that quarkyonic matter stiffens EOS, resulting in a higher mass. Conversely, the f-mode oscillation frequency decreases from 2.68 kHz to 1.89 kHz for G3 and 2.45 kHz to 2.08 kHz for IOPB-I at a transition density of $n_t=0.3 fm^{-3}$ and a confinement scale of $\Lambda_{cs} =800 $ MeV. Increasing the transition density from $0.3 fm^{-3}$ to $0.4 fm^{-3}$ and $0.5 fm^{-3}$  while keeping the confinement scale constant at $1400 $ MeV decreases the quark content, reflected in a lower mass and higher frequency. To investigate the effects of dark matter, we select a specific combination of  $n_t=0.3 fm^{-3}$ and $\Lambda_{cs} =800 $ MeV and examine its sensitivity to various DM compositions in panel (b). Increasing the DM percentage by varying $k_f^{DM}$ from 0.00 GeV (only quarkyonic matter) to 0.03 GeV and 0.04 GeV while maintaining the quark matter content has similar effects as seen in Panel (a). However, the resulting mass and frequency combinations differ significantly. We select the Fermi momentum of DM $k_f^{DM} =0.03 $ GeV case for a dark matter-admixed neutron star and compare it with pure baryonic and quarkyonic matter in panel (c). The DM admixed NS predicts intermediate mass $M = 2.54 M_\odot$ and frequency $f = 2.348 $ kHz for IOPB-I  and $M = 2.34 M_\odot$, $f = 2.218 $ kHz values for G3 as observed in panel (c). 
\item  The compactness of a star is defined by ($C=\frac{M}{R}$) and the $Z_s$ is called the redshift parameter quantifies the gravitational redshift experienced by a photon as it travels radially from the surface of the star to infinity. It is connected with C by the following simple equation:
\begin{eqnarray}
Z_{s} = (1-2C)^{\frac{1}{2}} -1
\label{eq:redshift}
\end{eqnarray}
By measuring the $Z_s$, one can directly constrain the compactness of the star, as well as other stellar properties such as mass, radius, and tidal deformability. Various studies in the literature have been conducted to estimate these stellar properties \cite{Lindblom}. In Fig. \ref{fig5} we have plotted the f mode frequency as a function of C for G3 (left panel) and IOPB-I (right panel) parameter sets. To see the effects of DM we have taken a quarkyonic star with $n_t = 0.4$ and $\Lambda = 1400 MeV$ configuration. Increasing the DM content essentially have very little effect on the compactness of the star with higher $f-$mode frequency. The relationship between f and $Z_s$ is depicted in Fig. \ref{fig2} for both G3 (upper panel) and IOPB-I (lower panel) parameter sets. To see the effects of the confinement scale $\Lambda_{cs}$, we fixed the $n_t = 0.3  \rm fm^{-3}$ and varied $\Lambda_{cs}$ from 800 MeV to 1400 MeV in panel (a). Both of the cases predict similar f mode frequency ($f_{max}$) and redshift ($Z_{s(max)}$) for the maximum mass of the star which is (2.080 kHz, 0,45) and (2.09 kHz, 0.49) for IOPB-I parameter set. Similar behaviour ensues for G3 also predicting a little higher frequency as well as redshift for a higher confinement scale, which is due to the slow rise in quark components in the quarkyonic matter. The effects of transition density are shown in panel (b) where the $\Lambda_{cs}$ is fixed to a value 1400 MeV while the transition density takes $0.3$  $\rm fm^{-3}$, $0.4$ $\rm  fm^{-3}$, and $0.5$ $\rm  fm^{-3}$, respectively. The results are significantly different from the previous scenario. As we move towards higher transition density, the $f_{max}$ increases and values of the redshift $Z_{s(max)}$ decreases. This behaviour can be attributed to the rapid decline in the quark component in quarkyonic matter. For example,  the ($f_{max}$, $Z_{s(max)}$) are (2.094 kHz, 0.49), (2.337 kHz, 0,48) and (2.418 kHz, 0.47) for $0.3$ $\rm fm^{-3}$, $0.4$ $\rm fm^{-3}$  and $ 0.5$ $\rm fm^{-3}$ respectively for IOPB-I parameter set. This behaviour is also exhibited by the G3 parameter set as well. 
\item One of the key observables that can provide insights into the microscopic equation of state (EOS) of neutron stars is tidal deformability. During the inspiral phase, the intense gravitational interaction between two neutron stars causes them to deform, with the extent of this deformation depending on the EOS. The dimensionless tidal deformability is connected to the compactness (C) and the second tidal Love number (k$_2$) via a specific relationship.  \begin{eqnarray}
\Lambda = \frac{2 k_2}{3C^5}.
\label{eq:effm_total}
\end{eqnarray} 
To estimate the influence of $k_f^{DM}$ and $n_t$., the crucial parameter that affects f-mode frequencies and tidal deformability, we have shown the contour plot between the  $f_{1.4}$ and $k_f^{DM}$ (panel (a)) for a specific quarkyonic star ($n_t =0.3  $ $\rm fm^{-3}$ , $\Lambda_{cs} =1400 $ MeV )and $f_{1.4}$ and $n_t$  for a fixed percentage of DM ($k_f^{DM} =0.03$ GeV)(panel (b)) where the colour bar represents tidal deformability $\Lambda_{1.4}$ of the canonical star for G3 and IOPB-I parameter sets. It is observed that in the first scenario, the lighter shaded region (yellow) corresponds to higher tidal deformability region $\Lambda_{1.4} = 1050-1200$ with small $k_f^{DM}$ and small $f_{1.4}$. As the DM momenta increases and reaches up to $k_f^{DM} = 0.04$ GeV from 0.00 GeV, we move towards a darker shaded region (deep purple). This region corresponds to a lower tidal deformability region with a higher percentage of DM and higher $f_{1.4}$. In panel (b), the higher tidal deformability region has small $f_{1.4}$ and small $n_t$. As the $n_t$ increases from 0.30 to 0.50, we move towards the darker shaded region which corresponds to a higher lower tidal deformability region with higher $n_t$ and higher $f_{1.4}$.  For example, with $n_t = (0.3, 0.4, 0.5)$ ($fm^{-3}$) the value of $\Lambda_{1.4} = 350, 600, 750 (MeV)$ for the case of IOPB-I.
\end{itemize}    

\subsection{ Universal relations} \label{rb}

\begin{itemize}
\item  It is well established that the frequency of the f mode correlates with the average stellar density $\sqrt{\frac{M}{R^3}}$. This relationship can be described by considering the connection between the speed of sound and the time taken for the fluid perturbation to propagate inside the star. Andersson and Kokkotas formulated an empirical relation for the f-mode frequency by examining stellar models with different realistic equations of state (EOSs). Although their study did not consider EOSs involving quark matter, they discovered that the f-mode frequencies they derived adhered closely to this empirical formula, showing little dependence on the particular EOS employed. These types of EOS-independent relations are referred to as universal relations \cite{PhysRevD.83.024014, PhysRevD.89.044006, PhysRevC.99.045806}. In Fig. \ref{fig3}, the relationship between f-mode frequency and average density is shown. We have plotted all the combinations of quarkyonic admixed DMscenarios for G3 and IOPB-I parameters set along with the baryonic part. Now, we will derive the empirical relation across the full range of uncertainties in the parameter space, specifically $n_t$, $\Lambda_{cs}$, $k_f^{DM}$, and NM parameters. We see a linear relationship between f mode frequency and the average density of the star, the linear fitting will lead to an approximate relation
\begin{eqnarray}
f (\rm kHz) = 0.6282 + 2.0476 \sqrt{\frac{\bar M}{\bar R^3}},
\label{eq:effm_total}
\end{eqnarray}
which has a correlation coefficient of about $\sim 95\%$. Here we have the dimesionless variables defined as $\bar M = \frac{M}{1.4 M\odot}$ and $\bar R = \frac{R}{10 \rm km}$. 
\item In Fig. \ref{fig4}, we have plotted the scaled f-mode frequency $\omega M$ as a function of compactness (C) for all the combinations of DMadmixed quarkyonic star along with the baryonic case. We observe a more consistent linear relationship between the two quantities that are reflected in the $99\%$ of the correlation coefficient, the fitting gives an approximate relation as:
\begin{eqnarray}
\omega M (\rm kHz \rm km) = -4.665 + 199.95 C,
\label{eq:effm_total}
\end{eqnarray}
\item We have shown the correlation heat map among various NS macroscopic properties like $M_{max}$, $R_{max}$, $R_{1.4}$, C, $\Lambda_{1.4}$, $I_{1.4}$, $f_{max}$, $f_{1.4}$ in Fig. \ref{fig7} for a specific DM admixed quarkyonic star ($n_t =0.3$ $\rm fm^{-3}$, $\Lambda_{cs} =1400 $ MeV , $k_f^{DM} = 0.03 \rm $ GeV). The correlations between $f_{1.4}$ – $\Lambda_{1.4}$, and $f_{max}$ – $\Lambda_{1.4}$ are shown in Fig. \ref{fig8}. We have taken all the 26 EOS that includes the baryonic, quarkyonic as well as DM admixed quarkyonic cases. This type of correlation is crucial as a measurement of tidal deformability from GW  data can lead to the constraint on the f-mode frequency of NS. 
\end{itemize}     
\section{Conclusions} 
\label{Conclusions}
In conclusion, we explored the $f-$mode frequencies of quarkyonic stars by integrating the effects of dark matter within the RMF formalism. This methodology enables us to analyze how DM influences the $f-$mode oscillation frequencies of NSs. Our calculations involves  using the relativistic Cowling approximation with two nuclear parameter sets: G3 and IOPB-I. The model comprises three free parameters: transition density, QCD confinement scale, and DM Fermi momentum, all of which have a significant impact on the $f-$mode oscillation properties. \\
In our study, we specifically considered the neutralino as a viable DM candidate, interacting with nucleons via a Yukawa potential mediated by Standard Model Higgs exchange. Notably, we found that dark matter-admixed quarkyonic stars conform to universal relations for f-mode frequencies, exhibiting a strong correlation coefficient. It is important to note that other oscillation modes in NSs, such as $f-$, $p-$, $g-$, and $w-$ modes, are currently undetectable by existing gravitational wave observatories like LIGO and VIRGO. Typically, f-mode oscillation frequencies range from 1-5 kHz, a domain that dark matter-admixed quarkyonic stars also occupy. Future studies may constrain these oscillation frequencies, providing crucial insights into the constituents of dense matter in such stellar objects.
\section{Acknowledgments}
One of the authors (JAP) is thankful to the Institute of Physics,  Bhubaneswar, for providing the computer facilities.
\bibliography{fmode}
\bibliographystyle{apsrev4-2}
\end{document}